\documentstyle[amssymb,aps]{revtex}

\begin{document}

\voffset=1.0cm 
\twocolumn[\hsize\textwidth\columnwidth\hsize
\csname @twocolumnfalse\endcsname

\title{A protocole to preserve quantum coherence in cavity QED}
\author{J. C. Retamal$^a$ and E. S. Guerra$^b$}
\address{$^a$Departamento de F\'{\i }sica, Universidad de Santiago
de Chile, Casilla 307 Correo 2, Santiago, Chile \\
$^b$Departamento de F\'{\i }sica, Universidade Federal rural do Rio de
Janeiro,\\ Caixa Postal 23851, seropedica, 23890 000, Rio de Janeiro,
Brazil}
\maketitle

\begin{abstract}
We consider the problem of quantum decoherence in cavity QED devices, and 
investigate the possibility to preserve a Schr\"odinger cat as a coherent
superposition along the time.  
\end{abstract}
] \narrowtext

A problem which has deserved a lot of attention in the present status of
quantum optics research is the investigation of the irreversible decay of
coherence in open systems\cite{Zurek1}. This is a fundamental problem, and
has many implications in present developments of quantum mechanical devices
like for instance cavity QED \cite{Davidovich}, trapped ions\cite{Monroe}
and quantum computers \cite{unruh}. As well known the irreversible loss of
quantum coherence in quantum optical systems is due to dissipation caused by
the unavoidable interaction with the surroundings of the system. Besides the
energy loss it has been well established that quantum coherence is depleted
as a function of time\cite{Zurek1}\cite{Haroche2}. The investigation of this
important issue is related with the possibility of generating
Schr\"{o}dinger cat states (macroscopic superpositions)\cite{Schrodinger}.
Schr\"{o}dinger cats of the vibrational motion of the center of mass of
trapped ions have been recently reported\cite{Monroe}. In cavity QED a
Schr\"{o}dinger cat generated via the superposition of coherent states of a
quantum field mode in a cavity has been proposed in a series of works\cite
{Nicim}. Recently, the monitoring of quantum coherence in a cavity QED
experiment \cite{Davidovich}\cite{Haroche2}, has established clear evidences
of the effects of quantum dissipation on a Schr\"{o}dinger cat. Other ideas
considering possible reversible behavior of quantum coherence in cavity QED
have been proposed \cite{Haroche3}.

The question we would like to address in this work is very simple: can a
Schr\"{o}dinger cat be preserved as a Schr\"{o}dinger cat along the time
even in the presence of dissipation? In this work we discuss this question
considering a quantized field in a microwave cavity in a quantum
superposition of two coherent states. Our approach is based on well known
proposals involving dispersive interactions between atomic systems and
quantized fields \cite{Nicim}. We will show that an appropriate preparation
of an atomic system which interacts dispersively with a dissipative
microwave cavity containing a Schr\"{o}dinger cat, can lead to an evolution
of the cavity field in which the coherence can be recovered. This is
achieved by a continuous probe of the atoms after the interaction with the
cavity field.

Typically dissipation in a microwave cavity occurs in a time scale of about $%
10^{-1}$ $s$. This time is very small as compared with the typical
interaction time associated with a Rydberg atom traveling through a cavity
at thermal velocities of around $300m/s$. At this velocities the interaction
time is about $0.1\mu s$. These times can be very small in comparison with
the typical decoherence time scale which is of the order of $\ t_{cav}/\bar{n%
}$. We assume that we use Rydberg atoms, because we can in practice neglect
atomic decay (as well as field dissipation) during the times involved in 
the protocol to be discussed below.

Present developments in cavity QED allows for the generation of well defined
initial conditions for the field state. In particular this is very
attractive in order to realize quantum mechanical experiments starting from
approximately the same initial conditions. Coherent states are a class of
reproducible initial field states. They allow for the generation, under
controlled experimental conditions, of the so called Schr\"odinger cat
states \cite{Haroche2}.

Let us consider a microwave cavity initially prepared with the field in a
Schr\"{o}dinger cat state 
\begin{equation}
\mid \psi \rangle =\frac{1}{\sqrt{2(1+e^{-2\mid \alpha \mid ^{2}})}}(\mid
\alpha \rangle +\mid -\alpha \rangle ).  \label{cat}
\end{equation}
Under the presence of dissipation at zero temperature the cavity dynamics is
described by \cite{scully} 
\begin{equation}
\dot{\rho}=\frac{\gamma }{2}(2a\rho a^{\dagger }-a^{\dagger }a\rho -\rho
a^{\dagger }a).  \label{disi1}
\end{equation}
where $\gamma $ is the cavity decay rate. From this equation it is not
difficult to show that the initial state given by Eq. (\ref{cat}) evolves to
the state 
\begin{equation}
\begin{array}[b]{ll}
\rho _{t}= & N_{+,0}^{2}[\mid \alpha _{t}\rangle \langle \alpha _{t}\mid
+\mid -\alpha _{t}\rangle \langle -\alpha _{t}\mid \\ 
& +(\mid \alpha _{t}\rangle \langle -\alpha _{t}\mid +\mid -\alpha
_{t}\rangle \langle \alpha _{t}\mid )e^{-2\mid \alpha \mid ^{2}(1-e^{-\gamma
t})})],
\end{array}
\label{rhocat1}
\end{equation}
where the normalization factor has been defined as $N_{+,0}=1/\sqrt{%
2(1+e^{-2\mid \alpha \mid ^{2}})}$ and we have defined $\alpha _{t}=\alpha
e^{-\gamma t/2}$. The irreversible decay of quantum coherence is manifestly
given by the time dependent factor in front of term $\mid -\alpha
_{t}\rangle \langle \alpha _{t}\mid $. It is very instructive to express
this result in the form 
\begin{equation}
\begin{array}{rr}
\rho _{t}= & \frac{1}{2}N_{+,0}^{2}\mid +,t\rangle \langle t,+\mid
(1+e^{-2\mid \alpha \mid ^{2}(1-e^{-\gamma t})}) \\ 
+ & \frac{1}{2}N_{+,0}^{2}\mid -,t\rangle \langle t,-\mid (1-e^{-2\mid
\alpha \mid ^{2}(1-e^{-\gamma t})})
\end{array}
\label{rhocat2}
\end{equation}
where we have introduced the definition 
\begin{equation}
\mid \pm ,t\rangle =\mid \alpha _{t}\rangle \pm \mid -\alpha _{t}\rangle .
\label{cat2}
\end{equation}
Notice that the states defined above are orthogonal, $\langle +,t$ $\mid
t,-\rangle =0$, but not normalized. This is very interesting because we have
written the density matrix in its diagonal form. This decomposition permits
us to envisage a possibility to project, after a measurement process, the
field state into one of the given orthogonal states. This measurement
process is the one involved in the standard protocol to generate
Schr\"{o}dinger cats. Here we apply the same scheme but from the point of
view of a quantum coherence preserving protocol.

Without any external influence, the field prepared in a Schr\"{o}dinger cat
state evolves to a statistical mixture, that is, $e^{-2\mid \alpha \mid
^{2}(1-e^{-\gamma t})}\rightarrow 0$ in Eq. (\ref{rhocat1}) for times of the
order of $t=1/(2\gamma \mid \alpha \mid ^{2})$. Taking into account Eq. (\ref
{rhocat2}) the decoherence process can be viewed as follows: the density
matrix expressed in its diagonal form has initially a nonvanishing
projection on the even cat, and as long as dissipation takes place, a
projection on the odd cat starts to build up, and then we have decoherence.

The dynamical behavior of quantum coherence occurs because of the
interaction with the reservoir and can not be avoided unless we can
interrupt and modify in some way the evolution of the system. In the problem
we are concerned here, the key idea is the keeping of the state of a system
by the projection into a pure state after a proper measurement. This could
be a trivial problem when considering quantum system with a few number of
states, but it could not be the case of a system with a large number of
states like a quantum field. In this sense we ask ourselves: can we find a
way of projecting dynamically a quantum field, which is evolving toward an
statistical mixture, into a pure state (for example a Schr\"{o}dinger cat)?

Let us consider a setup involving a Ramsey cavity $R_{1}$ followed by a
cavity $C$, and by another Ramsey cavity $R_{2}$ in a one dimensional array
along the $z$ axis. A set of detectors $D$ is set at the end of the path.
Assume that we have previously prepared a Schr\"{o}dinger cat in the cavity $%
C$ as in Eq. (\ref{cat}). Unavoidable cavity dynamics converts this initial
field into the state given by Eq. (\ref{rhocat2})). Now we let an atom to
travel along the $z$ axis to interact with the cavity. We assume that the
atom is conveniently prepared and its states are rotated in the Ramsey
cavities $R_{1}$and $R_{2}$ respectively, and finally probed in detector $D$%
. We will present two methods to achieve our main goal. We consider first an
effective two-level system as the projector element and alternatively we use
a three-level lambda atomic system in a degenerated configuration to play
the same role. The two schemes are just a different realization of the same
idea.

Consider a three-level cascade atom with $\mid i\rangle ,\mid e\rangle ,\mid
g\rangle $ being the upper, intermediate and lower atomic level. We assume
the transition $\mid i\rangle \rightarrow \mid e\rangle $ is detuned from
the cavity frequency. In addition we assume that $\mid e\rangle \rightarrow
\mid g\rangle $ transition is highly detuned from the cavity frequency and
is resonant with the Ramsey cavities frequency. We send an atom through the
cavities. First the atom is prepared in cavity $R_{1}$ in the superposition

\begin{equation}
\mid \psi \rangle =\frac{1}{\sqrt{2}}(\mid e\rangle +\mid g\rangle ).
\label{two1}
\end{equation}
Then the atom goes to cavity $C$ and interacts dispersively with the cavity
field and we can write the effective Hamiltonian 
\begin{equation}
H=\hbar \frac{g^{2}}{\Delta }a^{\dagger }a\mid e\rangle \langle e\mid +\mid
g\rangle \langle g\mid .  \label{H2}
\end{equation}
The global state for the atom-field state is expressed as 
\begin{equation}
\rho _{T}=\frac{1}{2}(e^{i\varphi a^{\dagger }a}\mid e\rangle \langle e\mid
+\mid g\rangle \langle g\mid )\rho _{t}(e^{-i\varphi a^{\dagger }a}\mid
e\rangle \langle e\mid +\mid g\rangle \langle g\mid ).
\end{equation}
Now let us consider the atom entering to a second Ramsey cavity $R_{2}$
where atomic levels are rotated according to the prescription 
\[
\mid e\rangle \rightarrow \frac{1}{\sqrt{2}}(\mid e\rangle +\mid g\rangle ), 
\]
\begin{equation}
\mid g\rangle \rightarrow \frac{1}{\sqrt{2}}(\mid e\rangle -\mid g\rangle ).
\label{tworota}
\end{equation}
Under this transformation, the state of the atom-field system changes to 
\begin{equation}
\begin{array}{rr}
\rho _{T}= & \Pi _{+}\rho _{t}\Pi _{+}^{\dagger }\mid e\rangle \langle e\mid
+\Pi _{+}\rho _{t}\Pi _{-}^{\dagger }\mid e\rangle \langle g\mid \\ 
+ & \Pi _{-}\rho _{t}\Pi _{+}^{\dagger }\mid g\rangle \langle e\mid +\Pi
_{-}\rho _{t}\Pi _{-}^{\dagger }\mid g\rangle \langle g\mid ,
\end{array}
\label{rhoT2}
\end{equation}
where 
\begin{equation}
\Pi _{\pm }=\frac{1}{2}(e^{i\varphi a^{\dagger }a}\pm 1).  \label{catpro}
\end{equation}
Some simple relations follow from the above definitions when adjusting the
parameters to fulfill the condition $\varphi =\pi $. In this case we have 
\begin{eqnarray*}
\Pi _{+} &\mid &\alpha _{t}\rangle =\frac{1}{2}\mid +,t\rangle , \\
\Pi _{-} &\mid &\alpha _{t}\rangle =-\frac{1}{2}\mid -,t\rangle ,
\end{eqnarray*}

\begin{eqnarray}
\Pi _{\pm } &\mid &\pm ,t\rangle =\pm \mid \pm ,t\rangle ,  \label{relations}
\\
\Pi _{\pm } &\mid &\mp ,t\rangle =0.  \nonumber
\end{eqnarray}
In a third step we probe the atomic levels and we have two possibilities:
the atom is detected in level $\mid e\rangle $ or $\mid g\rangle .$
Consequently the field is projected respectively into the states, 
\begin{equation}
\begin{array}{l}
\rho _{t,e}=\Pi _{+}\rho _{t}\Pi _{+}^{\dagger }, \\ 
\rho _{t,g}=\Pi _{-}\rho _{t}\Pi _{-}^{\dagger },
\end{array}
\label{detection}
\end{equation}
so that after detection of the atomic levels we have respectively two
possibilities,

\[
\begin{array}{l}
\rho _{t,e}=N_{+,t}\mid +,t\rangle \langle t,+\mid , \\ 
\rho _{t,g}=N_{-,t}\mid -,t\rangle \langle t,-\mid .
\end{array}
\]
The probability associated to each event is given by

\begin{equation}
\begin{array}{l}
P_{e}=N_{+,0}^{2}(1+e^{-2\mid \alpha \mid ^{2}(1-e^{-\gamma
t})})(1+e^{-2\mid \alpha \mid ^{2}e^{-\gamma t}}), \\ 
P_{g}=N_{+,0}^{2}(1-e^{-2\mid \alpha \mid ^{2}(1-e^{-\gamma
t})})(1-e^{-2\mid \alpha \mid ^{2}e^{-\gamma t}}).
\end{array}
\label{probtwo}
\end{equation}
We observe from these relations that the probability to detect the atom in
the excited level $\mid e\rangle $ is larger when compared with the
probability to detect the atom in the lower level $\mid g\rangle $. We have $%
P_{e}>P_{g}$ for all times when starting from an even cat. This last
assertion provides us a way to effectively keep the field state as a
Schr\"{o}dinger cat for a sequence of atoms interacting with the cavity
within a time between atoms short compared with the decoherence time.
Therefore, we effectively have a protocol to preserve quantum coherence.

For a short time between consecutive atoms, the probability to detect the
atom in the excited state is large so that we could evaluate the probability
that $N$ atoms be detected in the upper state. Let us assume that we drive
periodically the cavity field so that within the decoherence time many atoms
are sent into the cavity following the steps described above. If we define
the time between consecutive atoms as 
\[
\Delta t=\frac{t_{d}}{N}, 
\]
where the decoherence time is 
\[
t_{d}=\frac{1}{2N\gamma \mid \alpha \mid ^{2}}, 
\]
then after $N$ atoms, it is not difficult to show that the probability that
in $N$ events we detect atoms in the excited state is

\[
P_{e}=\prod\limits_{n=0}^{N}N_{+n}^{2}(1+e^{-2\mid \alpha _{n}\mid
^{2}(1-e^{-\gamma \Delta t})})(1+e^{-2\mid \alpha _{n}\mid ^{2}e^{-\gamma
\Delta t}}), 
\]
where 
\[
N_{+n}=\frac{1}{\sqrt{2(1+e^{-2\mid \alpha _{n}\mid ^{2}})}} 
\]
and 
\[
\alpha _{n}=\alpha e^{-n\gamma \Delta t/2}. 
\]
So, we conclude that for a large number of atoms sent within a time scale of
the order of the decoherence time ($\Delta t\lesssim t_{d}$), the
probability for detection in the upper state $P_{e}\rightarrow 1$. Even if
we wait a certain time without sending atoms, we always have an appreciable
possibility of projecting the field state into a Schr\"{o}dinger cat state,
however, the probability to project into a odd cat state increases.

It is interesting to point out that a similar protocol can be obtained by
using degenerate three-level lambda atoms in a far off resonant
configuration. Let us consider a three level $\Lambda $ atom with two
degenerated lower levels $\mid b\rangle $ and $\mid c\rangle $ interacting
dispersively with the cavity mode. It is not difficult to show that in this
regime the $\Lambda $ atom field dynamics is described by the evolution
operator \cite{guerra} 
\[
U_{d}=\left( 
\begin{array}{ll}
\Pi _{+} & \Pi _{-} \\ 
\Pi _{-} & \Pi _{+}
\end{array}
\right) 
\]
Where the operators $\Pi _{\pm }$ are defined in Eq. (\ref{catpro}). The
attractive feature of using a $\Lambda $ system is an economy of steps in
the global process leading to the structure given in Eq. (\ref{rhoT2}).
Assume the atom is sent initially in the lower level $\mid b\rangle $. Then,
before the interaction, the atom-field state is given by 
\[
\rho _{T}=\rho _{t}\mid b\rangle \langle b\mid .
\]
After the interaction the atom-field system evolves to 
\[
\begin{array}{rr}
\rho _{T}= & \,\Pi _{+}\rho _{t}\Pi _{+}\mid b\rangle \langle b\mid +\Pi
_{+}\rho _{t}\Pi _{-}\mid b\rangle \langle c\mid  \\ 
+ & \Pi _{-}\rho _{t}\Pi _{+}\mid c\rangle \langle b\mid +\Pi _{-}\rho
_{t}\Pi _{-}\mid c\rangle \langle c\mid .
\end{array}
\]
From these relations we observe that detecting the atom in level $\mid
b\rangle $ give us, 
\[
\rho _{t,b}=\mid +,t\rangle \langle t,+\mid ,
\]
and detecting the atom in level $\mid c\rangle $ we have, 
\[
\rho _{t,c}=\mid -,t\rangle \langle t,-\mid .
\]
The probabilities of detecting the atoms in $\mid b\rangle $ or $\mid
c\rangle $ are the same as defined in Eq. (\ref{probtwo}). The same
arguments for the effective two-level atom case apply to this case when
considering a sequence of atoms within a time scale short compared with the
dissipation and decoherence time. A disadvantage in using this scheme is
related to how experimentally to distinguish the two closely lower levels
and initially prepare properly the atom. This is an experimental point we do
not address in this work.

An interesting extension of the ideas presented above is the possibility of
manipulating the quantum states associated to two or more correlated
cavities. In fact it is not difficult to show that, for example, two
cavities prepared in a state

\[
\mid \psi \rangle =N_{+}(\mid \alpha \rangle \mid \beta \rangle +\mid
-\alpha \rangle \mid -\beta \rangle ), 
\]
evolves under dissipation to

\begin{equation}
\begin{array}{rr}
\rho _{t}= & \frac{1}{2}N_{+}^{2}\mid +,t\rangle \langle t,+\mid
(1+e^{-2\mid \alpha \mid ^{2}(1-e^{-\gamma t})-2\mid \beta \mid
^{2}(1-e^{-\gamma t})}) \\ 
+ & \frac{1}{2}N_{+}^{2}\mid -,t\rangle \langle t,-\mid (1-e^{-2\mid \alpha
\mid ^{2}(1-e^{-\gamma t})-2\mid \beta \mid ^{2}(1-e^{-\gamma t})}),
\end{array}
\label{rhocorre}
\end{equation}
where the states $\mid \pm ,t\rangle $ are defined as

\[
\mid \pm ,t\rangle =(\mid \alpha _{t}\rangle \mid \beta _{t}\rangle \pm \mid
-\alpha _{t}\rangle \mid -\beta _{t}\rangle ).
\]
For simplicity we have assumed equal decay rates for the cavities. The
question now is: how the dispersive interaction of a three-level atom
described above with both cavities will influence the state of the cavities.
Assume that an atom initially prepared in a superposition $\mid \psi
_{0}\rangle =(\mid e\rangle +\mid g\rangle )/\sqrt{(}2)$ , crosses both
cavities and then the atomic states are rotated in a second Ramsey cavity.
This process leads to a final state, after the flight of the atom, which is
essentially equivalent to the one found in Eq. (\ref{rhoT2}). Then after
detection of the atomic state we find that the field state can be projected
into a state 
\[
\begin{array}{ll}
\rho _{t,e}= & N_{+,t}^{2}\mid +,t\rangle \langle t,+\mid , \\ 
\rho _{t,g}= & N_{+,t}^{2}\mid -,t\rangle \langle t,-\mid .
\end{array}
\]
The same result can be extended for an arbitrary number of coupled cavities.
Of course in this case there are additional limitations because of the
decreasing of the decoherence time scale which now is of the order of $%
t_{cav}/(\mid \alpha \mid ^{2}+\mid \beta \mid ^{2})$. In this case the beam
of atoms necessary to keep the correlated state in a coherent superposition
should have a shorter interval of time between consecutive atoms.

Naturally this conclusion is conditioned by the regularity of the atom
sequence, and the number of events that in practice can be detected. For
example, a reasonable question we can address is to consider what happens if
we do not detect one of the atoms in the sequence. Assume that for a
regularly spaced sequence of atoms we have detected always the atoms in the
upper state, until the $n$-th atom. Assume that the atom $n+1$ is not
detected. This means that the state after this atom has passed through the
cavity is

\begin{equation}
\rho _{n}=\Pi _{+}\rho _{n}\Pi _{+}^{\dagger }+\Pi _{-}\rho _{n}\Pi
_{-}^{\dagger },
\end{equation}
where $\rho _{n}$ is given by Eq (\ref{rhocat2}) for $\alpha _{t}=\alpha
e^{-n\gamma \Delta t}$ and $t=\Delta t$. If the atom is not detected the
initial condition now is given by

\begin{equation}
\begin{array}{rr}
\rho _{n+1}= & \frac{1}{2}N_{n}^{2}\mid +,n+1\rangle \langle n+1,+\mid
(1+e^{-2\mid \alpha _{n}\mid ^{2}(1-e^{-\gamma \Delta t})}) \\ 
+ & \frac{1}{2}N_{n}^{2}\mid -,n+1\rangle \langle n+1,-\mid (1-e^{-2\mid
\alpha _{n}\mid ^{2}(1-e^{-\gamma \Delta t})}).
\end{array}
\end{equation}
Subsequently the field evolves during the interval $\Delta t$ until the $n+2$
atom enters the cavity and afterwards it changes to the state 
\begin{equation}
\begin{array}{rr}
\rho _{n+2}= & \frac{1}{2}N_{n}^{2}\mid +,n+2\rangle \langle n+2,+\mid
(1+e^{-2\mid \alpha _{n}\mid ^{2}(1-e^{-2\gamma \Delta t})}) \\ 
+ & \frac{1}{2}N_{n}^{2}\mid -,n+2\rangle \langle n+2,-\mid (1-e^{-2\mid
\alpha _{n}\mid ^{2}(1-e^{-2\gamma \Delta t})}),
\end{array}
\label{rhocat5}
\end{equation}
so that a non detected atoms increase the probability of detecting the atom
in level $\mid g\rangle $%
\begin{equation}
\begin{array}{l}
P_{e}=N_{n}^{2}(1+e^{-2\mid \alpha _{n+2}\mid ^{2}(1-e^{-\gamma
t})})(1+e^{-2\mid \alpha _{n}\mid ^{2}e^{-\gamma t}}) \\ 
P_{g}=N_{n}^{2}(1-e^{-2\mid \alpha _{n+2}\mid ^{2}(1-e^{-\gamma
t})})(1-e^{-2\mid \alpha _{n}\mid ^{2}e^{-\gamma t}}).
\end{array}
\end{equation}
The non detected atoms disturb the possibility of keeping the state of the
field as a Schr\"{o}dinger cat state. However, for short enough time between
consecutive atoms, the upper level probability is still large even if we do
not detect some of the atoms.

An interesting question which can be envisaged is the following: can be
there another kind of superposition of field states that lead to a density
matrix which under dissipation is kept in a diagonal form? It is reasonable
to think that the answer to these question could be yes. However, there
seems to be something special in the even and odd cat state basis which lead
a density matrix in a diagonal form under dissipation. We note that this
states are a natural decomposition of the all Hilbert space into subspaces
involving only even and only odd number states.

Finally we point out that the search of a mechanism combining a protocol
like the above one and a mechanism to preserve energy could keep the
coherence of a state of a fixed finite energy, avoiding the natural
evolution of a coherent superposition towards the vacuum. This is a crucial
goal in cavity QED.

Acknowledgments: One of us (J. R.) aknowledge to Dora, Moises and Catalina
for their support. Finnancial support from Fondecyt 1990838 and Dicyt
(USACH) is gratefully acknowledged..

\end{document}